\documentclass[a4paper,12pt]{article}
\usepackage[utf8]{inputenc}
\usepackage{graphicx}
\usepackage{amssymb,amsmath,amsthm}
\usepackage{subfigure}
\usepackage[margin=2cm]{geometry}
\usepackage{tikz}
\usepackage{authblk}
\usetikzlibrary{calc,patterns,angles,quotes} 
\usepackage{url}

\title{Computing on actin bundles network}
\author[1]{Andrew Adamatzky}
\affil[1]{Unconventional Computing Laboratory, Department of Computer Science, University of the West of England, Bristol, UK}
\author[2]{Florian Huber}
\affil[2]{Netherlands eScience Center, Science Park 140, 1098 XG Amsterdam, The Netherlands}
\author[3]{J\"{o}rg Schnau{\ss}}
\affil[3]{Soft Matter Physics Division, Peter Debye Institute for Soft Matter Physics, Faculty of Physics and Earth Science, Leipzig University, Germany \& Fraunhofer Institute for Cell Therapy and Immunology (IZI), DNA Nanodevices Group, Leipzig, Germany}
\date{}

\date{}

\graphicspath{{figs/}}

\begin{document}

\maketitle

\begin{abstract}
\noindent
Actin filaments are conductive to ionic currents, mechanical and voltage solitons. These travelling localisations can be utilised in making the actin network executing specific computing circuits. The propagation of localisations on a single actin filament is experimentally unfeasible, therefore we propose a `relaxed' version of the computing on actin networks by considering excitation waves propagating on actin bundles. We show that by using an arbitrary arrangement of electrodes it is possible to implement two-inputs-one-output circuits. Frequencies of the Boolean gates' detection in actin network match an overall distribution of gates discovered in living substrates. 

\vspace{2mm}

\end{abstract}

\section{Introduction}

An idea of implementation of computation by using collisions of signals travelling along one-dimensional non-linear geometries can be traced back to mid 1960s when Atrubin developed a chain of finite-state machines executing multiplication~\cite{atrubin1965one}, Fisher designed prime numbers generators in cellular automata~\cite{fischer1965generation} and Waksman proposed the eight-state solution for a firing squad synchronisation problem~\cite{waksman1966optimum}. 
In 1986, Park, Steiglitz, and Thurston~\cite{park1986soliton} designed a parity filter cellular automata with soliton-like dynamics of localisation. Their design led to a construction of a 1D particle machine, which performs computation by colliding particles in 1D cellular automata, i.e. computing embedded in a bulk media~\cite{squier1994programmable}. Being inspired by translating the purely theoretical ideas of collision-based computing~\cite{badamatzky,adamatzkyCBC} into nano-computing at the subcellular level, we consider actin and tubuline networks being ideal candidates for the computing substrates. The idea of subcellular computing on cytoskeleton networks has been firstly announced by Hameroff and Rasmussen in a context of microtubule automata in 1980s~\cite{hameroff1989information,rasmussen1990computational,hameroff1990microtubule}.
Also, Priel, Tuszynski and Cantiello analysed how information processing could be realised in actin-tubulin networks of neuron dendrites~\cite{priel2006dendritic}. In the present paper we focus on actin.

Actin is a crucial protein presented in all eukaryotic cells in forms of monomeric, globular actin (G-actin) and filamentous actin (F-actin)~\cite{straub1943actin, korn1982actin, szent2004early}. Under the appropriate conditions, G-actin polymerises into filamentous actin forming a double helical structure~\cite{huber2013advances}. Within a filament, actin monomers display slightly changed shapes compared to their free, globular configuration~\cite{oda2009nature}. Signals in actin networks could be represented by travelling localisations. The existence of the travelling localisations --- defects, ionic waves, solitons --- in cytoskeleton polymer networks is supported by (bio)-physical models~\cite{tuszynski1995ferroelectric,tuszynski2004results,tuszynski2004ionic, tuszynski2005molecular, priel2006ionic, tuszynski2005nonlinear,sataric2009nonlinear, sataric2010solitonic,sataric2011ionic, kavitha2017localized}.

Computational studies demonstrated that it is feasible to consider implementing Boolean gates on a single actin filament~\cite{siccardi2016boolean} and on an intersection of several actin filaments~\cite{siccardi2016logical} via collisions between solitons and to use a reservoir-computing-like approach to discover functions on a single actin unit~\cite{adamatzky2017logical} and filament~\cite{adamatzky2018discovering}. In 2016, for instance, we demonstrated that it is possible to implement logical circuits by linking the protein chains~\cite{siccardi2016logical}. In such a setup, Boolean values are represented by localisations travelling along the filaments and computation is realised via collisions between localisations at the junctions between the chains. We have shown that {\sc and}, {\sc or} and {\sc not} gates can be implemented in such setups. These gates can be cascaded into hierarchical circuits, as we have shown on an example of {\sc nor}~\cite{siccardi2016logical}. 

The theoretical models developed so far address processing of information on a single actin unit or a chain of few units. Whilst being attractive from a computing point, the model might be difficult to implement in experimental laboratory conditions. Therefore, we developed an alternative version of the computing on actin networks by considering excitation waves propagating on actin bundles. Not a single actin filament is considered but an overall `density' of the conductive material formed by the actin bundles arranged by crowding effects without the need of additional accessory proteins~\cite{schnaussPRL2016,schnaussreview2016}. First results of the innovative approached are presented below.

\section{Model}

FitzHugh-Nagumo (FHN) equations~\cite{fitzhugh1961impulses,nagumo1962active,pertsov1993spiral} give us a qualitative approximation of the Hodgkin-Huxley model~\cite{beeler1977reconstruction} of electrical activity of living cells:
\begin{eqnarray}
\frac{\partial v}{\partial t} & = & c_1 u (u-a) (1-u) - c_2 u v + I + D_u \nabla^2 \\
\frac{\partial v}{\partial t} & = & b (u - v),
\end{eqnarray}
where $u$ is a value of a trans-membrane potential, $v$ a variable accountable for a total slow ionic current, or a recovery variable responsible for a slow negative feedback, $I$ a value of an external stimulation current. The current through intra-cellular spaces is approximated by
$D_u \nabla^2$, where $D_u$ is a conductance. Detailed explanations of the `mechanics' of the model are provided in~\cite{rogers1994collocation}, here we shortly repeat some insights. The term $D_u \nabla^2 u$ governs a passive spread of the current. The terms $c_2 u (u-a) (1-u)$ and $b (u - v)$ describe the ionic currents. The term $u (u-a) (1-u)$ has two stable fixed points $u=0$ and $u=1$ and one unstable point $u=a$, where $a$ is a threshold of an excitation.

We integrated the system using the Euler method with the five-node Laplace operator, a time step $\Delta t=0.015$ and a grid point spacing $\Delta x = 2$, while other parameters were $D_u=1$, $a=0.13$, $b=0.013$, $c_1=0.26$. We controlled excitability of the medium by varying $c_2$ from 0.09 (fully excitable) to 0.013 (non excitable). Boundaries are considered to be impermeable: $\partial u/\partial \mathbf{n}=0$, where $\mathbf{n}$ is a vector normal to the boundary. 

\begin{figure}[!tbp]
    \centering
\subfigure[]{\includegraphics[width=0.49\textwidth]{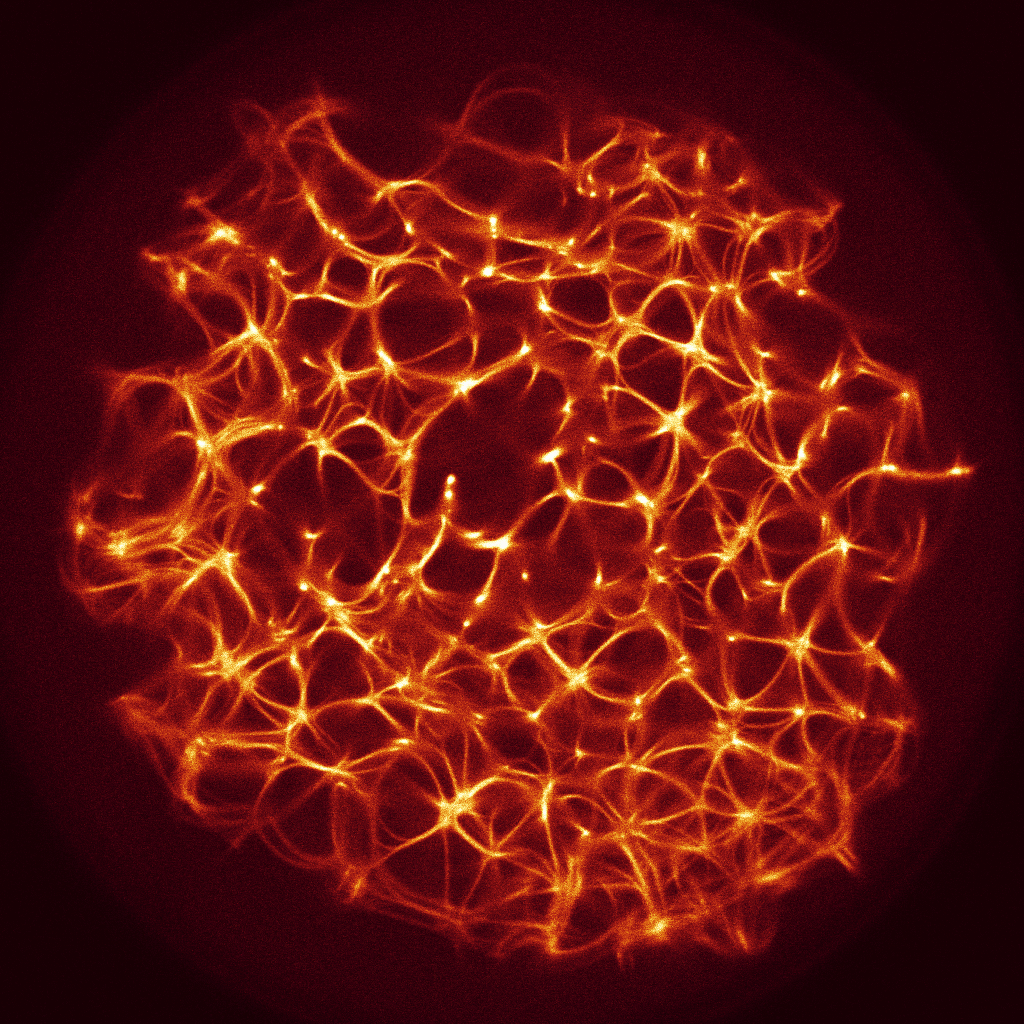}\label{originalimage}}
\subfigure[]{\includegraphics[width=0.49\textwidth]{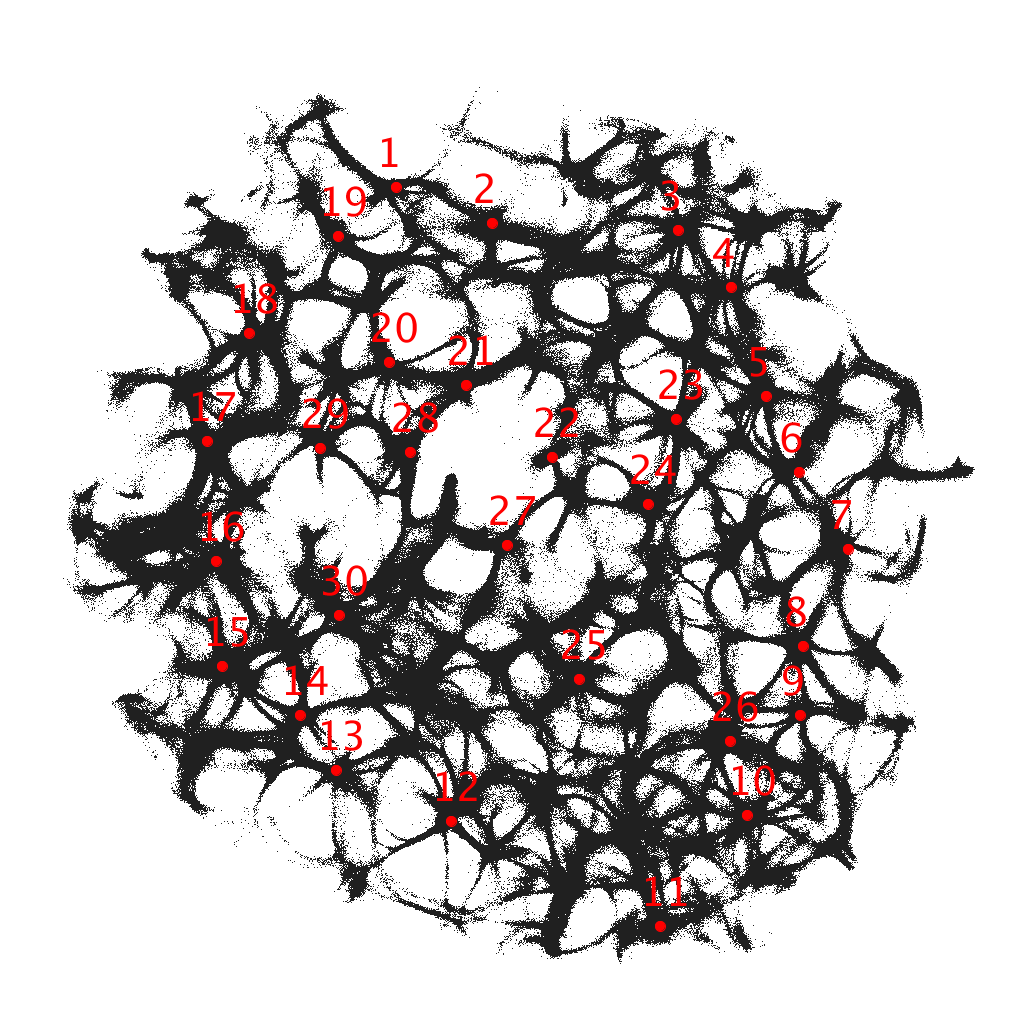}\label{matrix}}
    \caption{(a)~Original image, which was published in~\cite{huber2015formation}.   (b)~`Conductive' matrix selected from (a) Locations of electrodes $E_1 \ldots E_{30}$ are shown by their indexes.}
    \label{fig:dataImage}
\end{figure}
We used still images of the actin network, produced in laboratory experiments on formation of regularly spaced bundle networks from homogeneous filament solutions~\cite{huber2015formation} as a conductive template. We have chosen this particular because it results from an experimental protocol which reliably produces regularly spaced aster-based networks formed due to cross-linking and bundling mechanisms in the absence of molecular motor-driven processes or other accessory proteins~\cite{huber2015formation}. These structures effectively form very stable and long-living 3d networks, which can be readily imaged with confocal LSM and subsequently displayed as 2d structures. Thus, these networks can form a hardware of future cytoskeleton based computers~\cite{adamatzky2018towards}.

The actin network  (Fig.~\ref{originalimage}) was projected onto a $1024 \times 2014$ nodes grid.  Original image $M=(m_{ij})_{1 \leq i,j \leq n}$, $m_{ij} \in \{ r_{ij}, g_{ij}, b_{ij} \}$, where $n=1024$ and $1 \leq r, g, b \leq 255$ (Fig.~\ref{originalimage}) was converted to a conductive matrix $C=(m_{ij})_{1 \leq i,j \leq n}$ (Fig.~\ref{matrix}) derived from the image as follows: $m_{ij}=1$  if $r_{ij}>40$, $(g_{ij}>19)$ and $b_{ij}>19$. 

\begin{figure}[!tbp]
    \centering
\subfigure[$c_2=0.1$]{\includegraphics[width=0.32\textwidth]{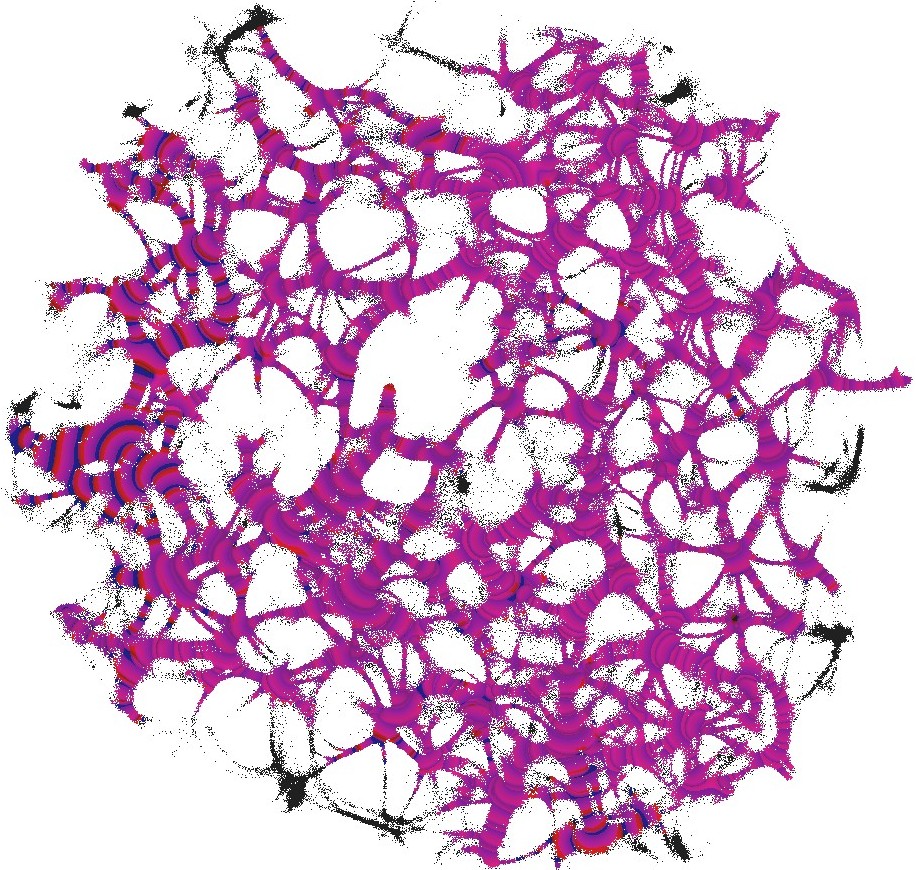}}
\subfigure[$c_2=0.105$]{\includegraphics[width=0.32\textwidth]{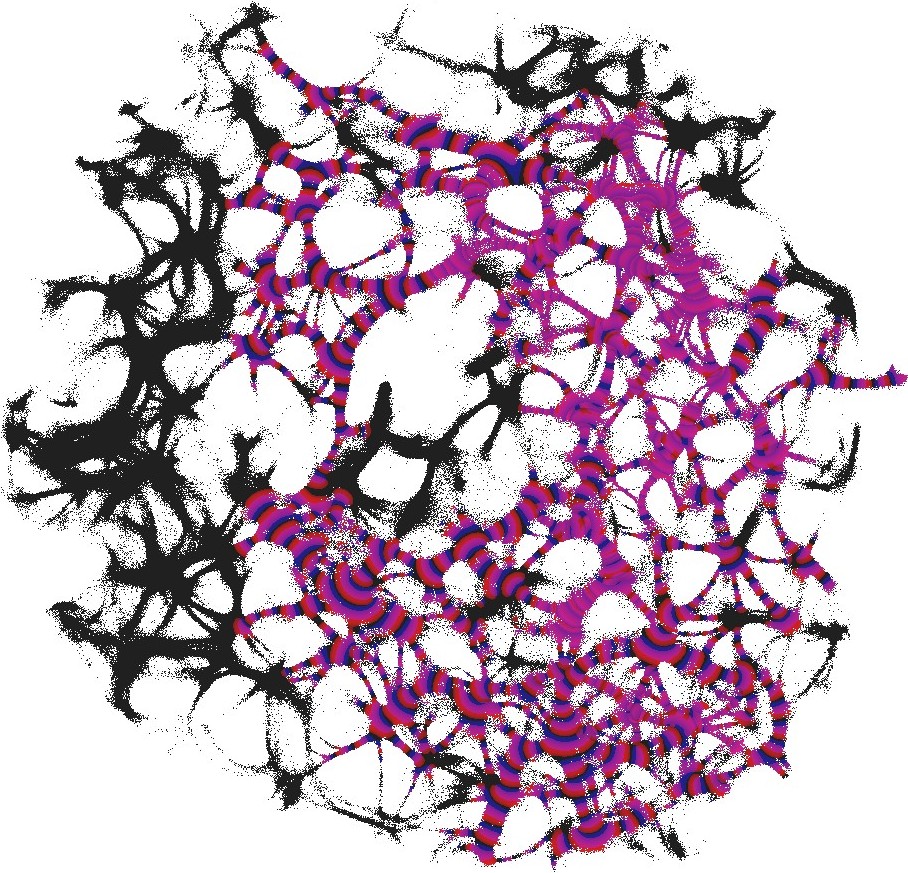}}
\subfigure[$c_2=0.106$]{\includegraphics[width=0.32\textwidth]{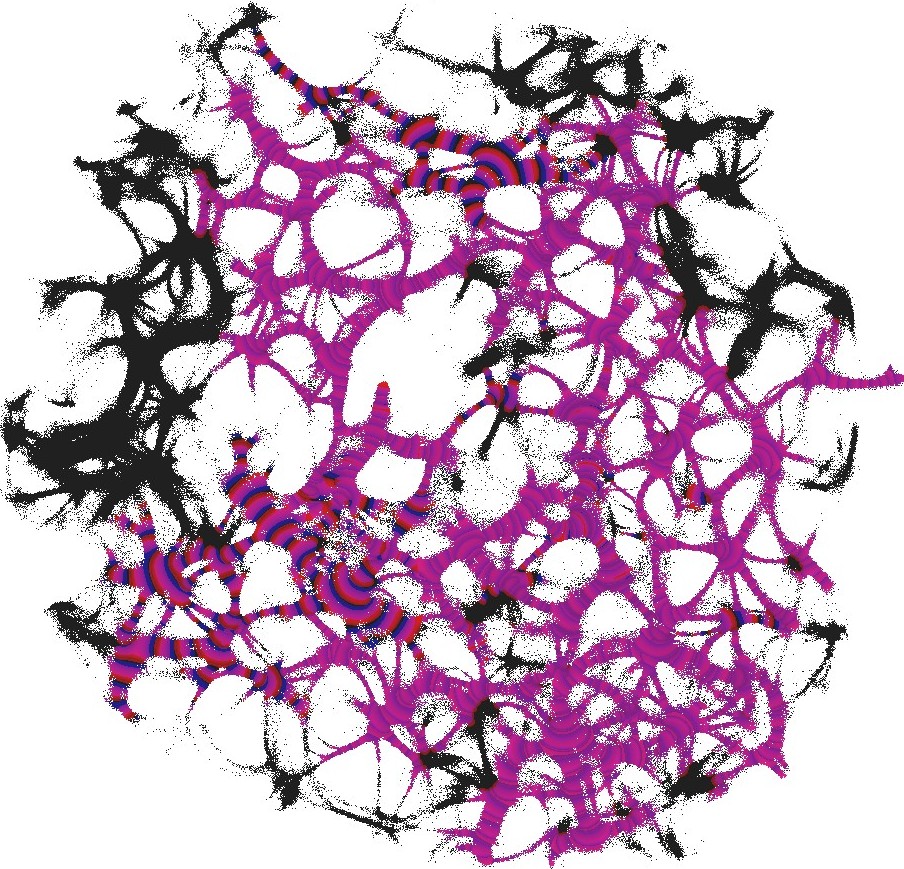}}
\subfigure[$c_2=0.1070$]{\includegraphics[width=0.32\textwidth]{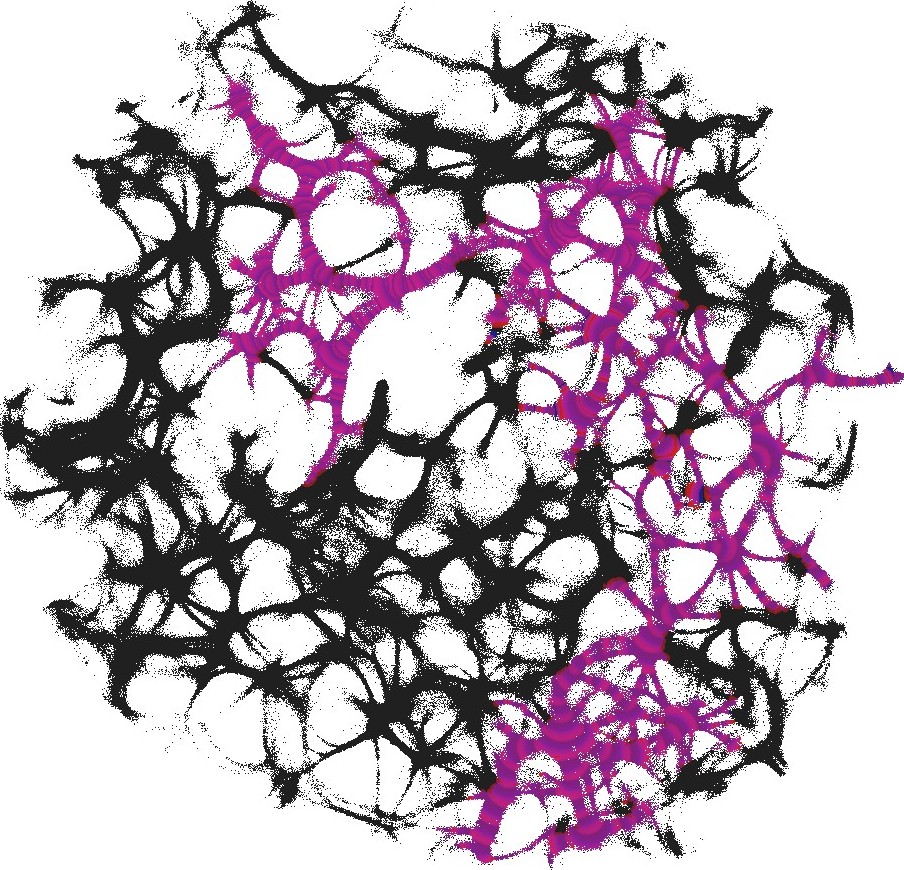}}
\subfigure[$c_2=0.1080$]{\includegraphics[width=0.32\textwidth]{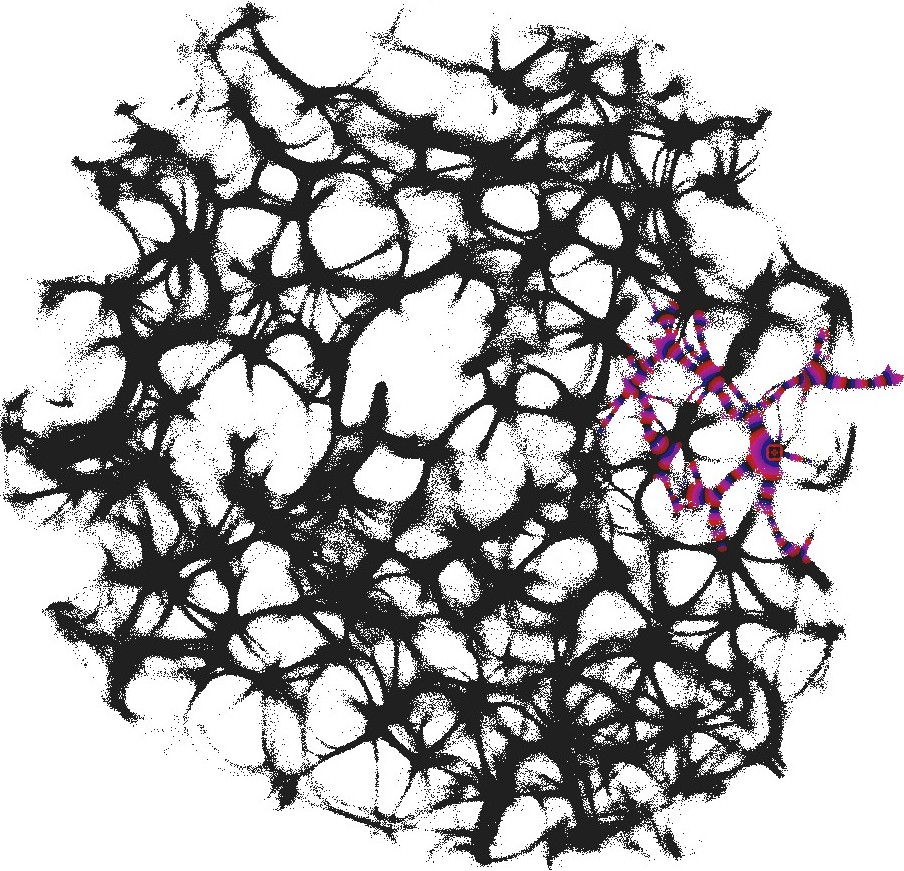}}
\subfigure[$c_2=0.1100$]{\includegraphics[width=0.32\textwidth]{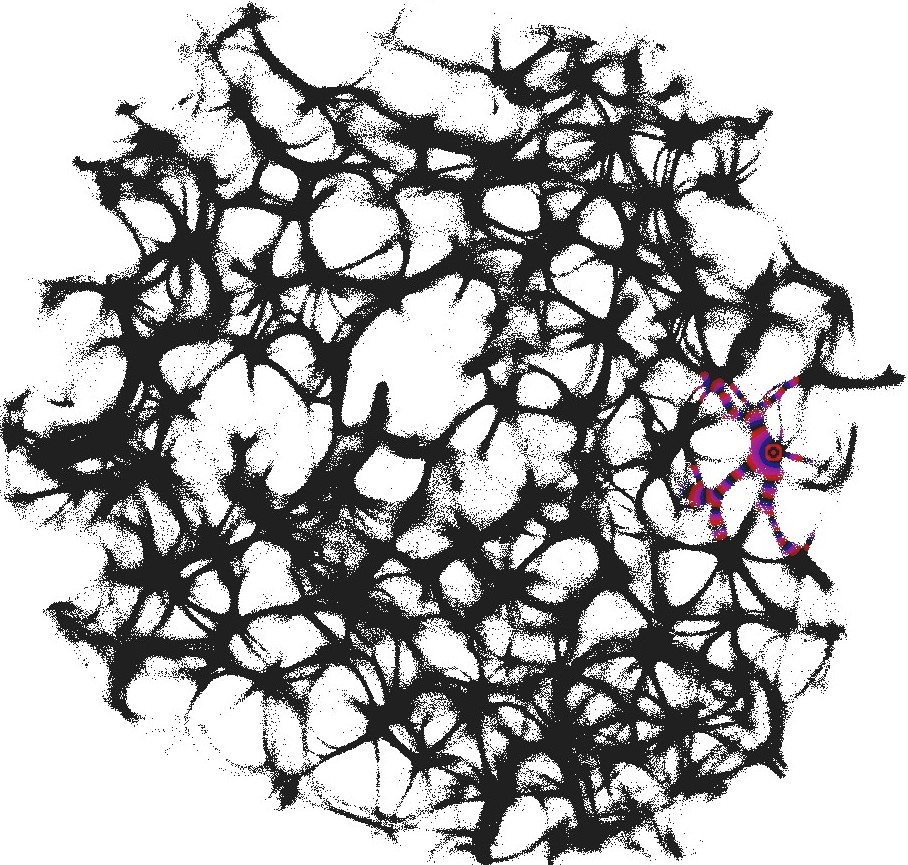}}
    \caption{Time lapse images of excitation wave-fronts propagating on the network displayed in Fig.~\ref{originalimage} for selected values of $c_2$. In each trial excitation was initiated at the site labelled `7' in Fig.~\ref{matrix}. Excitation wave-fronts are shown in red, conductive sites in black.}
    \label{fig:c2coverage}
\end{figure}

The parameter $c_2$ determines excitability of the medium and thus determines a range of the network coverage by excitation waves fronts. This is illustrated in Fig.~\ref{fig:c2coverage}.

To show dynamics of both $u$ and $v$, we calculated a potential $p^t_x$ at an electrode location $x$ as $p_x = \sum_{y: |x-y|<2} (u_x - v_x)$. Locations of electrodes $E_1, \cdots, E_{30}$ are shown in Fig.~\ref{matrix}. 

The numerical integration code written in Processing  was inspired by \cite{hammer2009, pertsov1993spiral,rogers1994collocation}.
Time-lapse snapshots provided in the paper were recorded at every 150\textsuperscript{th} time step, and we display sites with $u >0.04$; videos and figures were produced by saving a frame of the simulation every 100\textsuperscript{th} step of the numerical integration and assembling them in the video with a play rate of 30 fps. Videos are available at \url{https://zenodo.org/record/2561273}.

\section{Results}

Input Boolean values are encoded in excitation as follows. We earmark two sites of the network as dedicated inputs, $x$ and $y$, and represent logical {\sc True}, or `1', as an excitation. If $x=1$ then site corresponding to $x$ is excited, if $x=0$ the site is not excited. There are several ways to represent output values: presence/absence of an excitation wave-front at a dedicated output site, patterns of spike activity in the network and frequencies of the activity in dedicated output domains. We present four prototypes of logical: structural gate (exact timing of collisions between excitation wave-fronts is determined geometrically), frequency-based gate (Boolean values of outputs are encoded into frequencies of excitation), integral activity gates (an activity of the whole network is encoded into Boolean values) and spiking gates (where logical are values are represented by spikes or their combinations and search for the gates is done by using many output electrodes scattered in the network). 

\subsection{Structural gate}

\begin{figure}[!tbp]
    \centering
\subfigure[]{\includegraphics[width=0.25\textwidth]{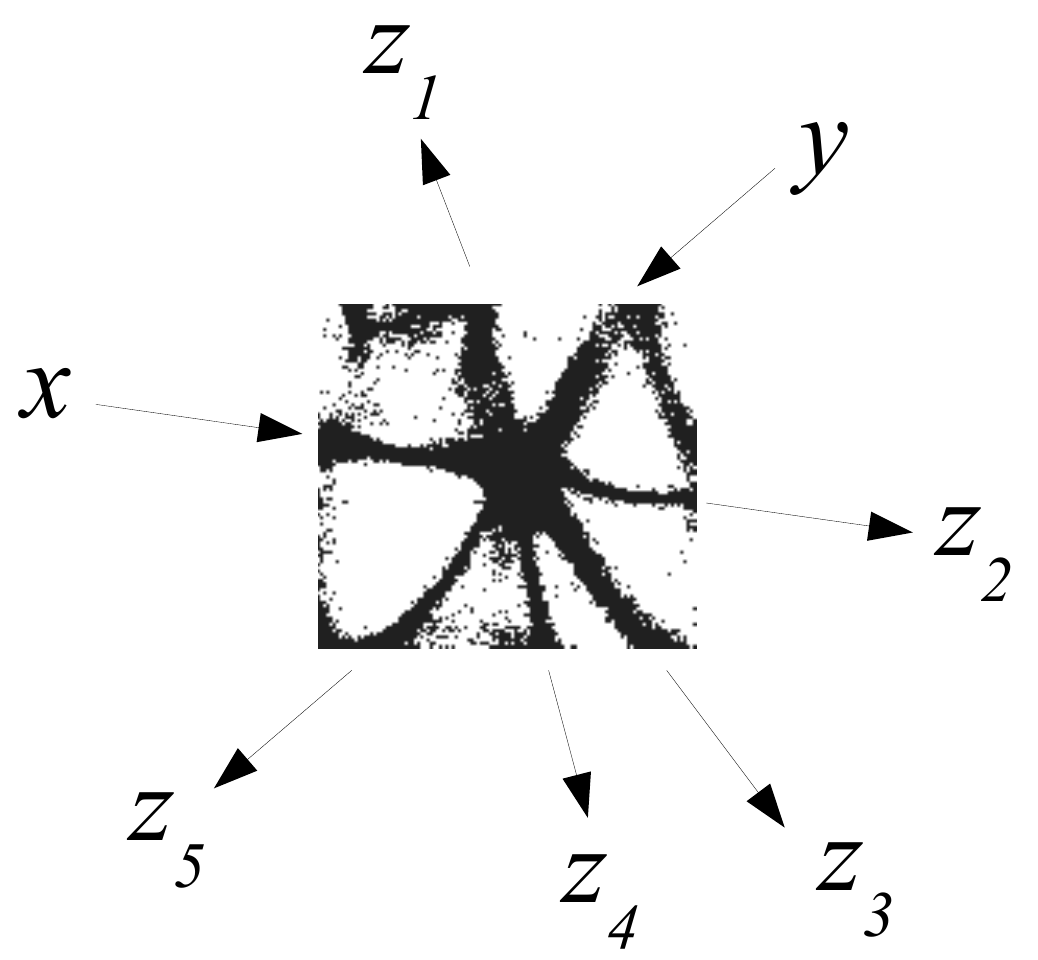}\label{gate1}}\\
    \subfigure[$x=0, y=1$]{\includegraphics[width=0.32\textwidth]{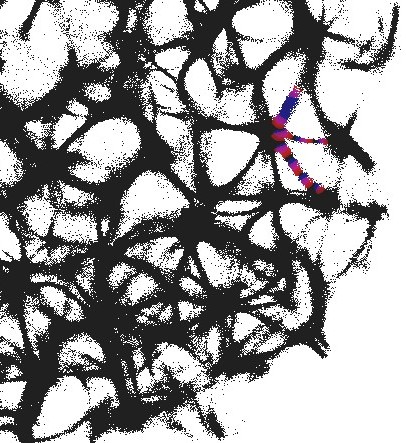}\label{gate01}}
    \subfigure[$x=1, y=0$]{\includegraphics[width=0.32\textwidth]{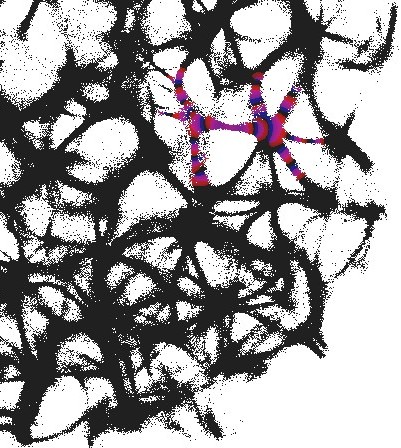}\label{gate10}}
    \subfigure[$x=1, y=1$]{\includegraphics[width=0.32\textwidth]{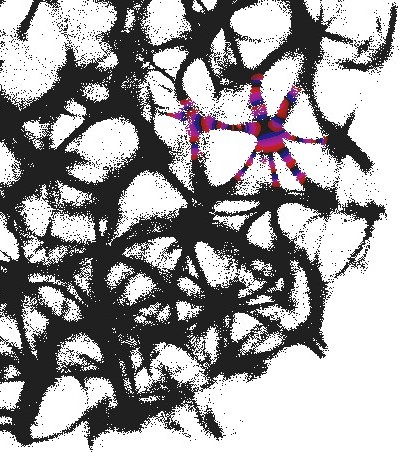}\label{gate11}}
    \caption{Interaction gate. (a)~A scheme of input and output channels. (b, c, d)~Time lapse images of wave-fronts propagating in the gate for inputs (b)~$x=0$ and $y=1$, (c)~$x=1$ and $y=0$, (d)~$x=1$ and $y=1$. Excitability of the medium is $c_2=0.108$}
    \label{fig:gateAndLapses}
\end{figure}

An example of an interaction gate  is shown in Fig.~\ref{fig:gateAndLapses}.  The gate is a junction of seven actin bundles, we call them `channels' (Fig.~\ref{gate1}). We earmark two channels as inputs $x$ and $y$, and five other channels as outputs $z_1, \ldots, z_5$.  To represent $x=1$ we excite channel $x$, to represent $y=1$ we excite channel $y$. When only channel $y$ is stimulated the excitation wave-fronts propagate into channels $z_2$ and $z_3$ (Fig.~\ref{gate01}). When only channel $x$ is stimulated, the excitation is recorded in channels $z_1$, $z_2$, $z_3$ (Fig.~\ref{gate10}). When both channels are excited, $x=1$ and $y=1$, the excitation propagates into all channels (Fig.~\ref{gate11}). Thus, the following functions are implemented on the output channels $z_1=x$, $z_2=z_3=x+y$, $z_4=z_5=xy$. The channel $z_1$ is a selector function. The channels $z_2$ and $z_3$ realise disjunction. The channels $z_4$ and $z_5$ implement conjuction. An advantage of the interaction gate is that it is cascadable, i.e. many gates can be linked together without decoders or couplers. A disadvantage is that functioning of the gate is determined by exact geometrical structure of the actin bundle network, which might be difficult to  control precisely.

\subsection{Frequency based gates}

\begin{figure}[!tbp]
    \centering
    \subfigure[]{\includegraphics[scale=0.3]{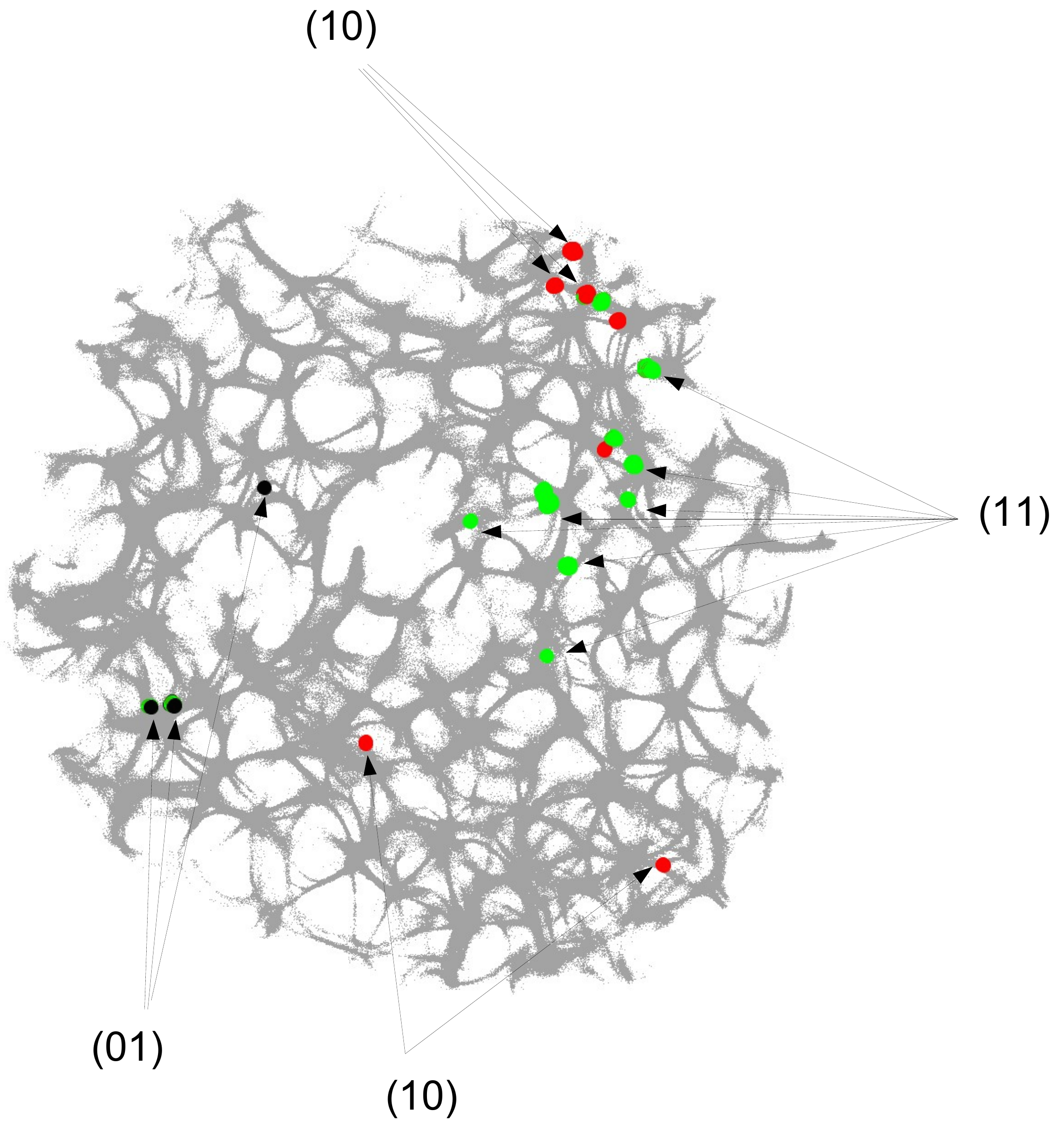}\label{frequencygatesc2_01}}
    \subfigure[]{\includegraphics[scale=0.3]{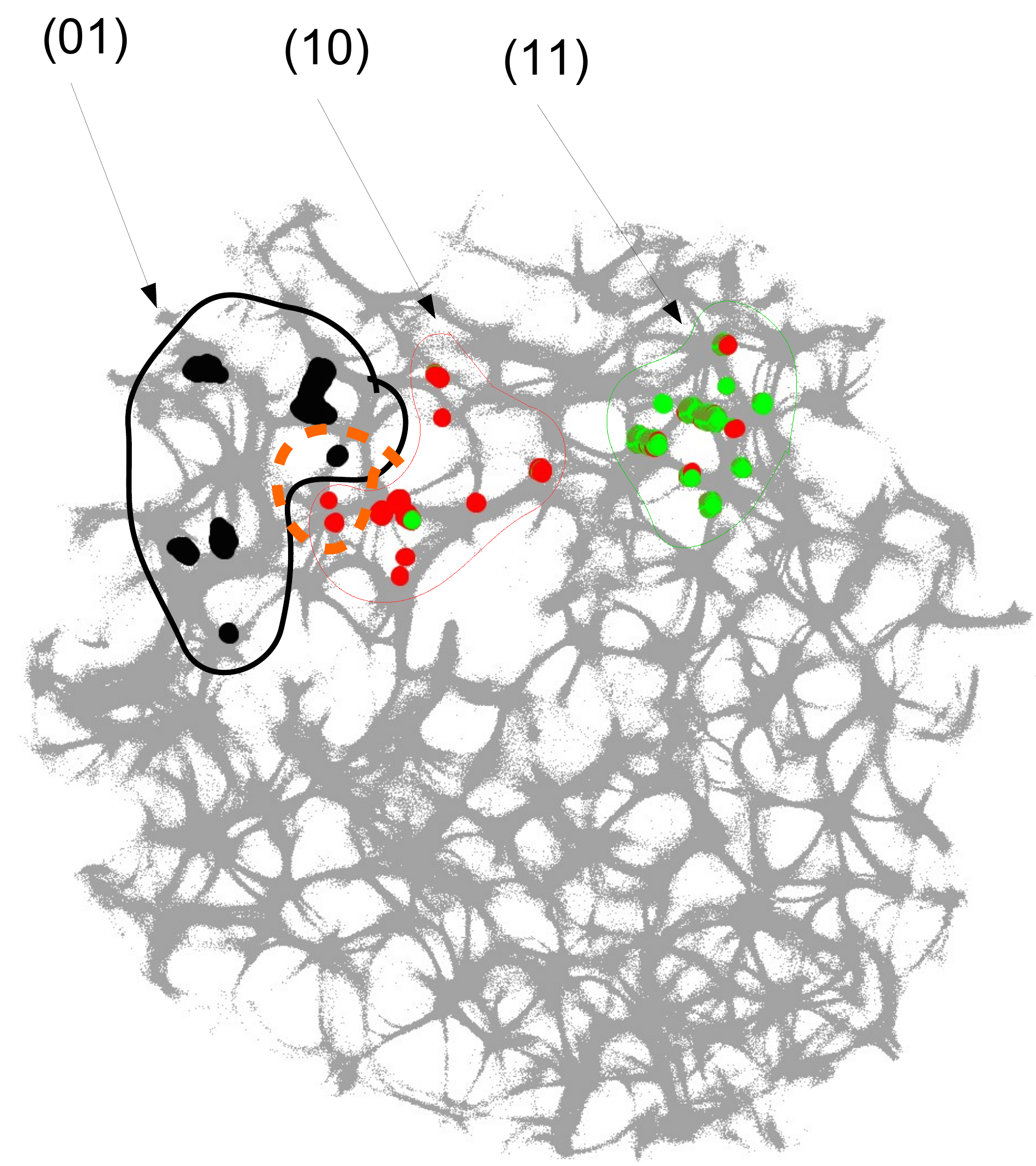}\label{frequencygatesc2_0107}}
    \caption{
    Domains with highest frequency of excitation represent  spatially separated outputs. They are shown by black discs for inputs $x=0$ and $y=1$, red discs for inputs $x=1$ and $y=0$ and green discs for inputs $x=1$ and $y=1$. Inputs $x$ and $y$ are sites $E_7$ and $E_{17}$ in (Fig.~\ref{matrix}).  
    (a)~$c_2=0.1$. (b)~$c_2=0.107$.}
    \label{fig:frequencytranversing}
\end{figure}

For each pair of inputs $(xy)$ $h \in \{01, 10, 11 \}$ 
we calculated a frequency matrix $\Omega_h = (\omega_{hs})$, $s \in \bf L$, where each entity with coordinates $s$ show how often a node $s$ of $\bf L$ was excited. At every iteration $t$ of the simulation  at every node $s$ we updated the frequency as follows: $\omega^t_s=\omega^t_s+1$ if $u^t_s>0.1$. When the simulation ends, the frequencies in all nodes were normalised as $\omega_s=\omega_s/\max\{\omega_z | z \in {\bf L}\}$. For each of $\Omega_h$  we selected domains of higher frequency as having entities $\omega_s>0.72$. These domains are shown in Fig.~\ref{fig:frequencytranversing}. This unique mapping allows to implement any two-input-one-output logical gates by placing electrodes in the required unique domains. For example, by placing electrodes in domains which represent outputs for both input pair $(01)$ and input pair $(10)$ (black  and red discs in Fig.~\ref{fig:frequencytranversing}), we can realise a {\sc xor} gate. 

While in excitable mode, $c_2=0.1$, domains corresponding to different input tuples are somewhat dispersed in the network (Fig.~\ref{frequencygatesc2_01}), the sub-excitable medium, $c_2=0.107$, shows compact and well spatially separated domains (Fig.~\ref{frequencygatesc2_0107}). Moreover, for $c_2=0.107$ we even observe a localised domain (shown by orange dashed contour in Fig.~\ref{frequencygatesc2_0107}), where input tuples $(01)$ and $(10)$ are displayed and thus the {\sc xor} gate is realised. 

\subsection{Overall level of activity}

\begin{figure}[!tbp]
    \centering
    \subfigure[]{\includegraphics[scale=0.45]{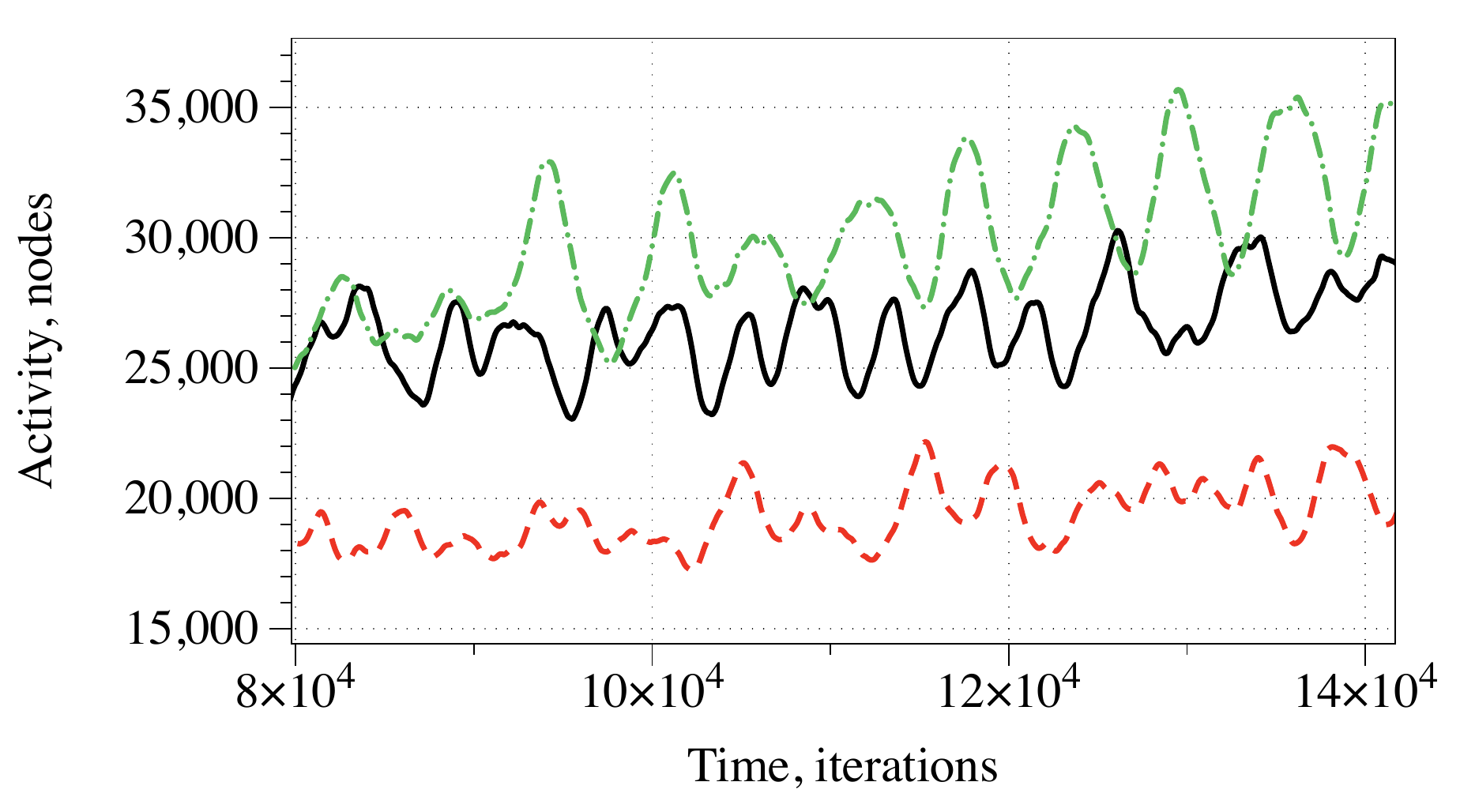}\label{activity_c2_01}}
    \subfigure[]{\includegraphics[scale=0.45]{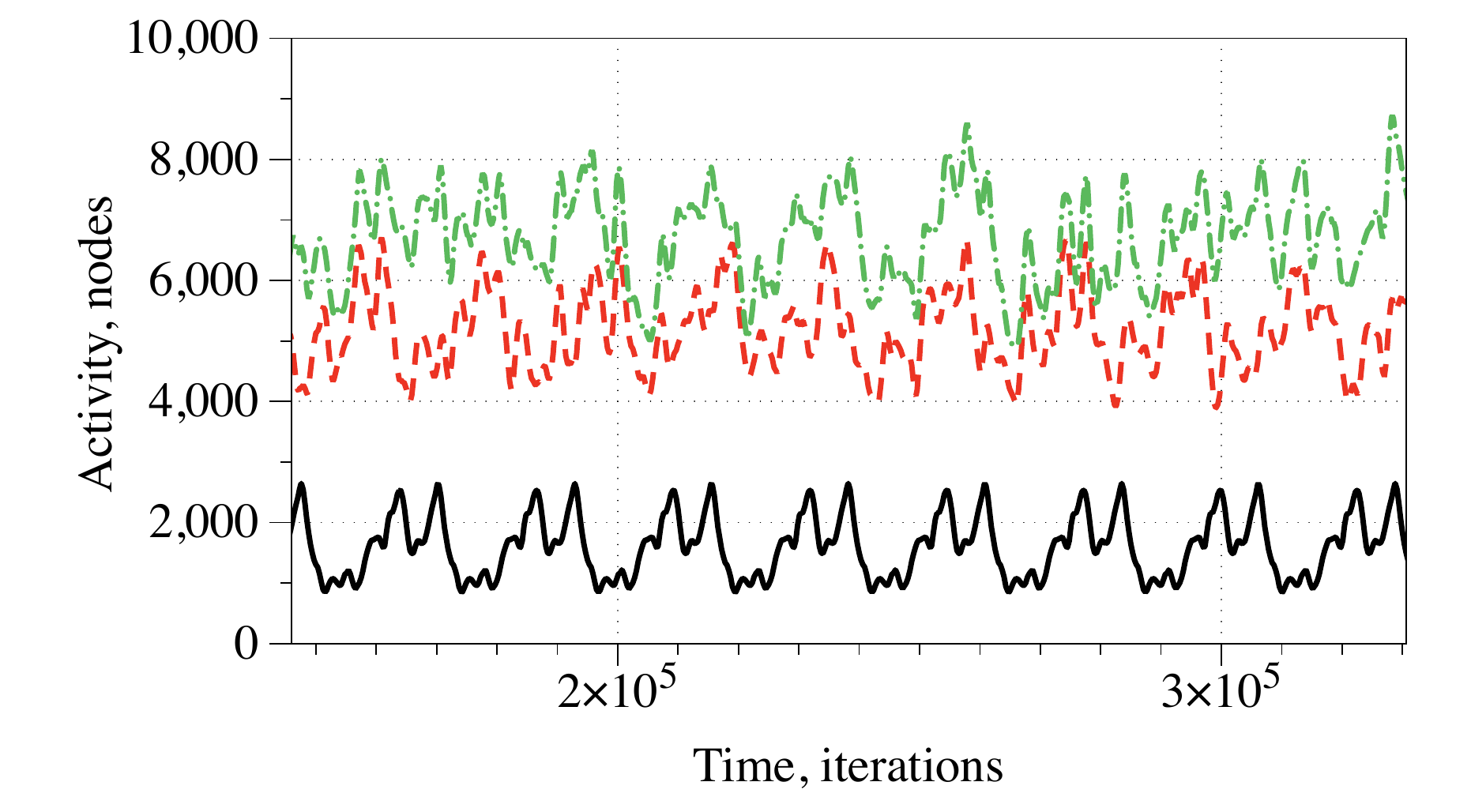}\label{activity_c2_0107}}
    \caption{Activity for input pairs $(xy)$ (stimulation sites are $E_7$ and $E_{17}$ in (Fig.~\ref{matrix})): (01) --- black solid, (10) --- red dashed, (11) --- dashed dotted.
    (a)~$c_2=0.1$. (b)~$c_2=0.107$.}
    \label{fig:activity}
\end{figure}

At every iteration $t$ we measured the activity of the network as a number of conductive nodes $x$ with $u^t_x>0.1$. A stimulation of the resting network evokes travelling wave-fronts, which collide with each other and may annihilate or form new wave-fronts in the result of the collisions. The wave-fronts also travel along cycling pathways in the network. Typically, e.g. after $8\times 10^4$ iterations for $c_2=0.1$ and $10^5$ iterations for $c_2=0.107$ the system falls in one of the limit cycle of the overall level of activity with a range of superimposed oscillations (Fig.~\ref{fig:activity}). We found no evidence that shapes of the superimposed spikes in activity reflect the exact combination of inputs, however, an average level of activity definitely does. A correspondence between input tuples and average level of activity $A$, as a percentage of a total number of conductive nodes is the following: 
$$
\begin{tabular}{c|ccc}
          &      & $(xy)$ &    \\
$c_2$     & (01) & (10) & (11) \\ \hline
0.1       & 0.068 & 0.05 & 0.08 \\
0.107     & 0.006 & 0.02 & 0.02 
\end{tabular}
$$
By  selecting an interval of $A'$ as a logical {\sc True} we can implement a range of gates. Consider the scenario $c_2=0.1$: $xy$ for $A'=[0.075,0.085]$, $x\overline{y}$ for $A'=[0.045,0.055]$, $\overline{x}y$ for $A'=[0.063,0.073]$, $x \oplus y$  $A'=[0.045,0.073]$. In the scenario $c_2=0.107$, a range of gates, implementable via assigning an activity interval to  
{\sc True}, is limited to $\overline{x}y$ for $A'=[0.005,0.007]$ and $x$ for $A'=[0.015,0.025]$.

\subsection{Spiking gate}  

\begin{figure}[!tbp]
    \centering
    \includegraphics[width=\textwidth]{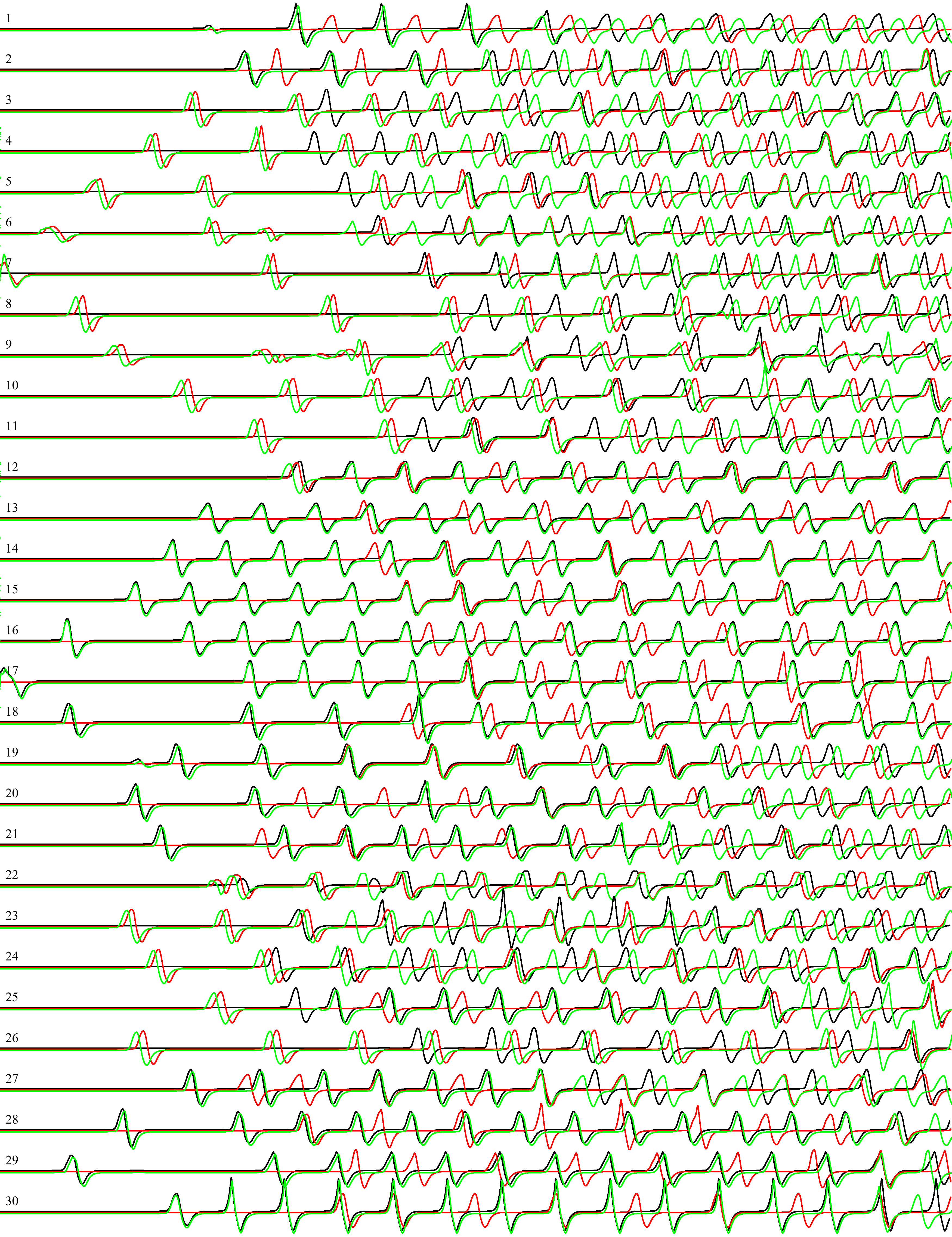}
    \caption{Potential recorded at 30 electrodes (Fig.~\ref{matrix}) during c. $14.2 \cdot 10^4$ iterations. Indexes of electrodes are shown on the left. Black lines show potential when the network was stimulated by input pair (01), red by (10) and green by (11). Excitability of the medium is $c_2=0.1$.}
    \label{fig:allSpikes_c2_0100}
\end{figure}

\begin{table}[!tbp]
    \centering
    \begin{tabular}{c|cc}
    spikes    & gate  & notations   \\  \hline
\includegraphics[scale=0.2]{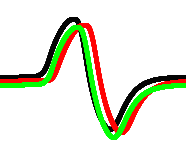}  & {\sc or} & $x+y$ \\
\includegraphics[scale=0.2]{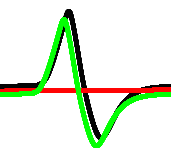}  & {\sc select} & $y$ \\
\includegraphics[scale=0.2]{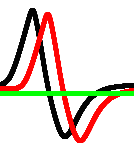}  & {\sc xor} & $x \oplus y$ \\
\includegraphics[scale=0.2]{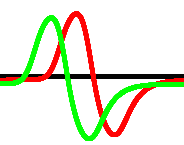}  & {\sc select} & $x$ \\
\includegraphics[scale=0.2]{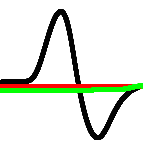}  & {\sc not-and} & $\overline{x}y$ \\
\includegraphics[scale=0.2]{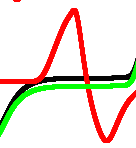}  & {\sc and-not} & $x\overline{y}$ \\
\includegraphics[scale=0.2]{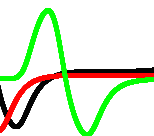}  & {\sc and } & $xy$ \\
    \end{tabular}
    \caption{Representation of gates by combinations of spikes. Black lines show the potential when the network was stimulated by input pair (01), red by (10) and green by (11).}
    \label{tab:spikes2gates}
\end{table}

\begin{figure}[!tbp]
    \centering
    \includegraphics[width=0.9\textwidth]{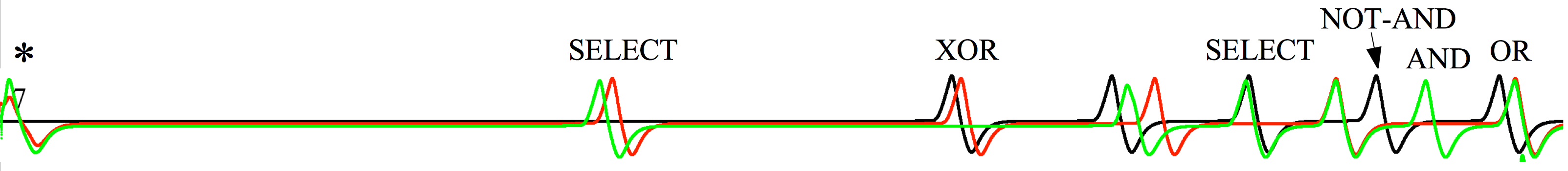}
    \caption{Spikes recorded at electrode $E_7$. Moment of initial stimulation is shown by *. Black lines show the potential when the network was stimulated by input pair (01), red by (10) and green by (11).}
    \label{fig:Spiking7}
\end{figure}

\begin{table}[!tbp]
    \centering
\begin{scriptsize}
    \begin{tabular}{c|cccccccc}
$i$ ($E_i$)	&	$x+y$	&	$y$	&	$x \oplus y$	&	$x$	&	$\overline{x}y$	&	$x\overline{y}$	&	$xy$	&	Total	\\ \hline
1	&	0	&	7	&	1	&	3	&	2	&	5	&	3	&	21	\\
2	&	0	&	8	&	1	&	3	&	2	&	5	&	3	&	22	\\
3	&	2	&	2	&	0	&	6	&	5	&	3	&	4	&	22	\\
4	&	2	&	3	&	0	&	6	&	5	&	3	&	3	&	22	\\
5	&	2	&	2	&	0	&	6	&	4	&	3	&	4	&	21	\\
6	&	1	&	4	&	1	&	5	&	4	&	2	&	5	&	22	\\
7	&	1	&	4	&	1	&	4	&	2	&	2	&	2	&	16	\\
8	&	0	&	3	&	1	&	6	&	3	&	1	&	2	&	16	\\
9	&	1	&	2	&	0	&	7	&	3	&	1	&	0	&	14	\\
10	&	1	&	2	&	0	&	8	&	4	&	0	&	0	&	15	\\
11	&	1	&	3	&	0	&	4	&	3	&	3	&	3	&	17	\\
12	&	2	&	7	&	0	&	3	&	0	&	3	&	0	&	15	\\
13	&	3	&	10	&	0	&	1	&	0	&	3	&	0	&	17	\\
14	&	2	&	11	&	0	&	1	&	0	&	4	&	0	&	18	\\
15	&	0	&	11	&	0	&	4	&	0	&	3	&	0	&	18	\\
16	&	0	&	11	&	0	&	3	&	1	&	4	&	0	&	19	\\
17	&	2	&	9	&	0	&	2	&	0	&	2	&	0	&	15	\\
18	&	3	&	8	&	0	&	2	&	0	&	2	&	0	&	15	\\
19	&	1	&	5	&	0	&	5	&	2	&	2	&	2	&	17	\\
20	&	2	&	8	&	1	&	1	&	1	&	5	&	3	&	21	\\
21	&	2	&	7	&	0	&	3	&	1	&	5	&	3	&	21	\\
22	&	2	&	4	&	0	&	7	&	2	&	0	&	3	&	18	\\
23	&	3	&	5	&	1	&	6	&	2	&	1	&	4	&	22	\\
24	&	1	&	5	&	1	&	7	&	2	&	1	&	4	&	21	\\
25	&	2	&	7	&	0	&	3	&	2	&	3	&	1	&	18	\\
26	&	2	&	2	&	0	&	6	&	5	&	2	&	1	&	18	\\
27	&	2	&	6	&	1	&	5	&	2	&	2	&	3	&	21	\\
28	&	1	&	10	&	0	&	3	&	0	&	5	&	0	&	19	\\
29	&	0	&	9	&	0	&	4	&	0	&	4	&	0	&	17	\\
30	&	2	&	9	&	0	&	3	&	0	&	3	&	1	&	18	\\ \hline
Average	&	1.43	&	6.13	&	0.30	&	4.23	&	1.90	&	2.73	&	1.80	&	18.53	\\
St dev	&	0.94	&	3.06	&	0.47	&	1.96	&	1.67	&	1.46	&	1.65	&	2.57	\\
Median	&	2.00	&	6.50	&	0.00	&	4.00	&	2.00	&	3.00	&	2.00	&	18.00\\
    \end{tabular}
 \end{scriptsize}
    \caption{Number of gates discovered for each recording site. Excitability of the medium is $c_2=0.1$. }
    \label{tab:numberOfgates}
\end{table}

 A spiking activity of the network shown in Fig.~\ref{originalimage}, with $c_2=0.1$ in a response to stimulation via electrodes $E_7$ and $E_{17}$ recorded from electrodes $E_1, \cdots, E_{30}$ is shown in Fig.~\ref{fig:allSpikes_c2_0100}.  Assume each spike represents logical {\sc True} and that spikes being less than  $2 \cdot 10^2$ iterations closer to each other happen at the same moment. Then a representation of gates by spikes and their combination will be as shown in Tab.~\ref{tab:spikes2gates}. 
 
 By selecting some specific intervals of recordings we can realise several gates in a single site of recording. In this particular case we assumed that spikes are separated if a difference between their occurrence time is more than  $10^3$ iterations. An example is shown in Fig.~\ref{fig:Spiking7}. 
 
 To estimate logical richness of the network, we calculated frequencies of logical gates' discoveries. For each of the recording sites we calculated a number of gates realised during $14.2 \cdot 10^4$ iterations (Tab.~\ref{tab:numberOfgates}). In terms of `frequency' of appearance of gates during the simulation, the gates can be arranged in the following hierarchy, from most frequently to least frequent:
 {\sc select}~$\rhd$~\{{\sc and-not}, {\sc not-and}\}~$\rhd$~{\sc and }~$\rhd$~{\sc or}~$\rhd$~{\sc xor}.  

 The model can realise two-inputs-two-outputs logical gates when we consider values of two recording electrodes at the same specified interval. For example, a one-bit half-adder: one output is {\sc and} and another output is {\sc xor}, and Toffoli gate: one output is {\sc select} and another {\sc xor}. 

\section{Discussion}

In numerical experiments we demonstrated that logical gates can be implemented in actin bundle networks by various ways of mapping excitation dynamics of the network onto output space. We illustrate the approach with detailed constructions of structural, frequency-based and overall activity based gates. We concluded our study with a comprehensive analysis of spiking gates, where we constructed a frequency of gates hierarchy. The frequency hierarchy of gates discovered {\sc select}~$\rhd$~\{{\sc and-not}, {\sc not-and}\}~$\rhd$~\{{\sc or},  {\sc and }\}~$\rhd$~{\sc xor} matches, with some variations, hierarchies of gates discovered in living slime mould~\cite{harding2018discovering}, living  plants~\cite{adamatzky2018computers}, Belousov-Zhabotinsky chemical medium~\cite{toth2008dynamic}. The gate {\sc select} is dominating because it reflects a reachability of the recording site from the stimulation site:  excitation from one electrode reaches a recording site, while an excitation originated from another electrode does not. The gates  {\sc and-not} and {\sc not-and} represent the scenario when an excitation wave-front propagating from one input site blocks, for instance by its refractory tail, the wave-front propagating from another input site. Gate {\sc and} symbolises the situation when wave-fronts originated on both input sites must meet up at some point of their travel to traverse areas with lower excitability, for example loci where a narrow channel enter sudden expansion. When excitation wave-fronts from both input sites can reach a recording side without annihilating each other, the site implements gate {\sc or}.  The gate {\sc xor} reflects the situation when wave-fronts originated at different input sites cancel each other. Modelling results obtained in the paper are encouraging: they show that a computation can be implemented in an actin bundles network by recording excitation dynamics of the network at few arbitrarily selected domains. Thus, we believe, our further tasks will aim at experimental laboratory implementation of the gates. Based on our previous experimental work described in~\cite{huber2015formation}, we can extend this approach by specifically biasing the architectural design of these networks. Actin in its natural environment has many accessory proteins such as crosslinkers, which directly impact the properties of the bundle structures~\cite{Strehle2011}. We have recently been able to mimic the properties of these naturally occurring accessory proteins with DNA-based, artificial constructs, which can alter the properties of actin structures in a programmable fashion~\cite{Lorenzcrosslinker2018}. The geometrical design of these constructs can be readily tuned by choosing different numbers of binding domains and by altering the underlying DNA template connecting them. These templates can be designed to favour specific binding angles and the number of bundles per junction. They would only act as a molecular precursor for the architecture of the actin system without interfering in the bundle formation and information transport themselves. With the ability to tune the properties of the junctions we gain control over the computing potential. Theoretically, it would be even possible to mix different types of structural proteins to allow parallel processing of information~\cite{Golde2018}.

\bibliographystyle{plain}
\bibliography{bibliography}

\begin{thebibliography}{10}

\bibitem{badamatzky}
A.~Adamatzky.
\newblock Collision-based computing in biopolymers and their automata models.
\newblock {\em International Journal of Modern Physics C}, 11:1321--1346, 2000.

\bibitem{adamatzkyCBC}
Andrew Adamatzky, editor.
\newblock {\em Collision-based Computing}.
\newblock Springer, 2002.

\bibitem{adamatzky2017logical}
Andrew Adamatzky.
\newblock Logical gates in actin monomer.
\newblock {\em Scientific reports}, 7(1):11755, 2017.

\bibitem{adamatzky2018discovering}
Andrew Adamatzky.
\newblock On discovering functions in actin filament automata.
\newblock {\em arXiv preprint arXiv:1807.06352}, 2018.

\bibitem{adamatzky2018computers}
Andrew Adamatzky, Simon Harding, Victor Erokhin, Richard Mayne, Nina Gizzie,
  Frantisek Balu{\v{s}}ka, Stefano Mancuso, and Georgios~Ch Sirakoulis.
\newblock Computers from plants we never made: Speculations.
\newblock In {\em Inspired by nature}, pages 357--387. Springer, 2018.

\bibitem{adamatzky2018towards}
Andrew Adamatzky, Jack Tuszynski, Joerg Pieper, Dan~V Nicolau, Rossalia
  Rinalndi, Georgios Sirakoulis, Victor Erokhin, Joerg Schnauss, and David~M
  Smith.
\newblock Towards cytoskeleton computers. {A} proposal.
\newblock In Andrew Adamatzky, Selim Akl, and Georgios Sirakoulis, editors,
  {\em From parallel to emergent computing}. CRC Group/Taylor \& Francis, 2019.

\bibitem{atrubin1965one}
AJ~Atrubin.
\newblock A one-dimensional real-time iterative multiplier.
\newblock {\em IEEE Transactions on Electronic Computers}, 3:394--399, 1965.

\bibitem{beeler1977reconstruction}
Go~W Beeler and H~Reuter.
\newblock Reconstruction of the action potential of ventricular myocardial
  fibres.
\newblock {\em The Journal of physiology}, 268(1):177--210, 1977.

\bibitem{fischer1965generation}
Patrick~C Fischer.
\newblock Generation of primes by a one-dimensional real-time iterative array.
\newblock {\em Journal of the ACM (JACM)}, 12(3):388--394, 1965.

\bibitem{fitzhugh1961impulses}
Richard FitzHugh.
\newblock Impulses and physiological states in theoretical models of nerve
  membrane.
\newblock {\em Biophysical journal}, 1(6):445--466, 1961.

\bibitem{Golde2018}
Tom Golde, Constantin Huster, Martin Glaser, Tina Händler, Harald Herrmann,
  Josef~A. Käs, and Jörg Schnauß.
\newblock Glassy dynamics in composite biopolymer networks.
\newblock {\em Soft Matter}, 14(39):7970--7978, 2018.

\bibitem{hameroff1990microtubule}
Stuart Hameroff and Steen Rasmussen.
\newblock Microtubule automata: Sub-neural information processing in biological
  neural networks, 1990.

\bibitem{hameroff1989information}
Stuart~R Hameroff and Steen Rasmussen.
\newblock Information processing in microtubules: Biomolecular automata and
  nanocomputers.
\newblock In {\em Molecular Electronics}, pages 243--257. Springer, 1989.

\bibitem{hammer2009}
Peter Hammer.
\newblock Spiral waves in monodomain reaction-diffusion model, 2009.

\bibitem{harding2018discovering}
Simon Harding, Jan Koutn{\'\i}k, J{\'u}rgen Schmidhuber, and Andrew Adamatzky.
\newblock Discovering boolean gates in slime mould.
\newblock In {\em Inspired by Nature}, pages 323--337. Springer, 2018.

\bibitem{huber2013advances}
Florian Huber, J{\"o}rg Schnau{\ss}, Susanne Rönicke, Philipp Rauch, Karla
  Müller, Claus Fütterer, and Josef~A. K{\"a}s.
\newblock Emergent complexity of the cytoskeleton: from single filaments to
  tissue.
\newblock {\em Advances in Physics}, 62(1):1--112, 2013.

\bibitem{huber2015formation}
Florian Huber, Dan Strehle, J{\"o}rg Schnau{\ss}, and Josef K{\"a}s.
\newblock Formation of regularly spaced networks as a general feature of actin
  bundle condensation by entropic forces.
\newblock {\em New Journal of Physics}, 17(4):043029, 2015.

\bibitem{kavitha2017localized}
L~Kavitha, E~Parasuraman, A~Muniyappan, D~Gopi, and S~Zdravkovi{\'c}.
\newblock Localized discrete breather modes in neuronal microtubules.
\newblock {\em Nonlinear Dynamics}, 88(3):2013--2033, 2017.

\bibitem{korn1982actin}
Edward~D Korn.
\newblock Actin polymerization and its regulation by proteins from nonmuscle
  cells.
\newblock {\em Physiological Reviews}, 62(2):672--737, 1982.

\bibitem{Lorenzcrosslinker2018}
Jessica~S. Lorenz, Jörg Schnauß, Martin Glaser, Martin Sajfutdinow, Carsten
  Schuldt, Josef~A. Käs, and David~M. Smith.
\newblock Synthetic transient crosslinks program the mechanics of soft,
  biopolymer‐based materials.
\newblock {\em Advanced Materials}, 30(13):1706092, 2018.

\bibitem{nagumo1962active}
Jinichi Nagumo, Suguru Arimoto, and Shuji Yoshizawa.
\newblock An active pulse transmission line simulating nerve axon.
\newblock {\em Proceedings of the IRE}, 50(10):2061--2070, 1962.

\bibitem{oda2009nature}
Toshiro Oda, Mitsusada Iwasa, Tomoki Aihara, Yuichiro Ma{\'e}da, and Akihiro
  Narita.
\newblock The nature of the globular-to fibrous-actin transition.
\newblock {\em Nature}, 457(7228):441--445, 2009.

\bibitem{park1986soliton}
James~K Park, Kenneth Steiglitz, and William~P Thurston.
\newblock Soliton-like behavior in automata.
\newblock {\em Physica D: Nonlinear Phenomena}, 19(3):423--432, 1986.

\bibitem{pertsov1993spiral}
Arkady~M Pertsov, Jorge~M Davidenko, Remy Salomonsz, William~T Baxter, and Jose
  Jalife.
\newblock Spiral waves of excitation underlie reentrant activity in isolated
  cardiac muscle.
\newblock {\em Circulation research}, 72(3):631--650, 1993.

\bibitem{priel2006ionic}
Avner Priel, Jack~A Tuszynski, and Horacio~F Cantiello.
\newblock Ionic waves propagation along the dendritic cytoskeleton as a
  signaling mechanism.
\newblock {\em Advances in Molecular and Cell Biology}, 37:163--180, 2006.

\bibitem{priel2006dendritic}
Avner Priel, Jack~A Tuszynski, and Horacion~F Cantiello.
\newblock The dendritic cytoskeleton as a computational device: an hypothesis.
\newblock In {\em The emerging physics of consciousness}, pages 293--325.
  Springer, 2006.

\bibitem{rasmussen1990computational}
Steen Rasmussen, Hasnain Karampurwala, Rajesh Vaidyanath, Klaus~S Jensen, and
  Stuart Hameroff.
\newblock Computational connectionism within neurons: A model of cytoskeletal
  automata subserving neural networks.
\newblock {\em Physica D: Nonlinear Phenomena}, 42(1-3):428--449, 1990.

\bibitem{rogers1994collocation}
Jack~M Rogers and Andrew~D McCulloch.
\newblock A collocation-{G}alerkin finite element model of cardiac action
  potential propagation.
\newblock {\em IEEE Transactions on Biomedical Engineering}, 41(8):743--757,
  1994.

\bibitem{sataric2009nonlinear}
MV~Satari{\'c}, DI~Ili{\'c}, N~Ralevi{\'c}, and Jack~Adam Tuszynski.
\newblock A nonlinear model of ionic wave propagation along microtubules.
\newblock {\em European biophysics journal}, 38(5):637--647, 2009.

\bibitem{sataric2011ionic}
MV~Satari{\'c} and BM~Satari{\'c}.
\newblock Ionic pulses along cytoskeletal protophilaments.
\newblock In {\em Journal of Physics: Conference Series}, volume 329, page
  012009. IOP Publishing, 2011.

\bibitem{sataric2010solitonic}
MV~Satari{\'c}, D~Sekuli{\'c}, and M~{\v{Z}}ivanov.
\newblock Solitonic ionic currents along microtubules.
\newblock {\em Journal of Computational and Theoretical Nanoscience},
  7(11):2281--2290, 2010.

\bibitem{schnaussPRL2016}
Jörg Schnauß, Tom Golde, Carsten Schuldt, B.~U.~Sebastian Schmidt, Martin
  Glaser, Dan Strehle, Tina Händler, Claus Heussinger, and Josef~A. Käs.
\newblock Transition from a linear to a harmonic potential in collective
  dynamics of a multifilament actin bundle.
\newblock {\em Phys. Rev. Lett.}, 116(10):108102, 2016.

\bibitem{schnaussreview2016}
Jörg Schnauß, Tina Händler, and Josef~A. Käs.
\newblock Semiflexible biopolymers in bundled arrangements.
\newblock {\em Polymers}, 8(8):274, 2016.

\bibitem{siccardi2016logical}
Stefano Siccardi and Andrew Adamatzky.
\newblock Logical gates implemented by solitons at the junctions between
  one-dimensional lattices.
\newblock {\em International Journal of Bifurcation and Chaos}, 26(06):1650107,
  2016.

\bibitem{siccardi2016boolean}
Stefano Siccardi, Jack~A Tuszynski, and Andrew Adamatzky.
\newblock Boolean gates on actin filaments.
\newblock {\em Physics Letters A}, 380(1-2):88--97, 2016.

\bibitem{squier1994programmable}
Richard~K Squier and Ken Steiglitz.
\newblock Programmable parallel arithmetic in cellular automata using a
  particle model.
\newblock {\em Complex systems}, 8(5):311--324, 1994.

\bibitem{straub1943actin}
FB~Straub.
\newblock Actin, ii.
\newblock {\em Stud. Inst. Med. Chem. Univ. Szeged}, 3:23--37, 1943.

\bibitem{Strehle2011}
Dan Strehle, Jörg Schnauß, Claus Heussinger, Jos{\'e} Alvarado, Mark Bathe,
  Josef~A. Käs, and Brian Gentry.
\newblock Transiently crosslinked f-actin bundles.
\newblock {\em European Biophysics Journal}, 40(1):93--101, 2011.

\bibitem{szent2004early}
Andrew~G Szent-Gy{\"o}rgyi.
\newblock The early history of the biochemistry of muscle contraction.
\newblock {\em The Journal of general physiology}, 123(6):631--641, 2004.

\bibitem{toth2008dynamic}
Rita Toth, Christopher Stone, Andrew Adamatzky, Ben de~Lacy~Costello, and Larry
  Bull.
\newblock Dynamic control and information processing in the
  belousov--zhabotinsky reaction using a coevolutionary algorithm.
\newblock {\em The Journal of chemical physics}, 129(18):184708, 2008.

\bibitem{tuszynski2005molecular}
JA~Tuszy{\'n}ski, JA~Brown, E~Crawford, EJ~Carpenter, MLA Nip, JM~Dixon, and
  MV~Satari{\'c}.
\newblock Molecular dynamics simulations of tubulin structure and calculations
  of electrostatic properties of microtubules.
\newblock {\em Mathematical and Computer Modelling}, 41(10):1055--1070, 2005.

\bibitem{tuszynski1995ferroelectric}
JA~Tuszy{\'n}ski, S~Hameroff, MV~Satari{\'c}, B~Trpisova, and MLA Nip.
\newblock Ferroelectric behavior in microtubule dipole lattices: implications
  for information processing, signaling and assembly/disassembly.
\newblock {\em Journal of Theoretical Biology}, 174(4):371--380, 1995.

\bibitem{tuszynski2004results}
JA~Tuszynski, T~Luchko, EJ~Carpenter, and E~Crawford.
\newblock Results of molecular dynamics computations of the structural and
  electrostatic properties of tubulin and their consequences for microtubules.
\newblock {\em Journal of Computational and Theoretical Nanoscience},
  1(4):392--397, 2004.

\bibitem{tuszynski2004ionic}
JA~Tuszy{\'n}ski, S~Portet, JM~Dixon, C~Luxford, and HF~Cantiello.
\newblock Ionic wave propagation along actin filaments.
\newblock {\em Biophysical journal}, 86(4):1890--1903, 2004.

\bibitem{tuszynski2005nonlinear}
Jack Tuszy{\'n}ski, St{\'e}phanie Portet, and John Dixon.
\newblock Nonlinear assembly kinetics and mechanical properties of biopolymers.
\newblock {\em Nonlinear Analysis: Theory, Methods \& Applications},
  63(5-7):915--925, 2005.

\bibitem{waksman1966optimum}
Abraham Waksman.
\newblock An optimum solution to the firing squad synchronization problem.
\newblock {\em Information and control}, 9(1):66--78, 1966.

\end{thebibliography}

\section*{Author contributions statements} 

A.A., F.H., J.S.  undertook the research and wrote the manuscript. 

\section*{Competing interests}  
The authors declare they have no competing interests.

\end{document}